\documentclass[12pt]{article}
\usepackage{graphics,cite,amssymb,epsfig,float}
\usepackage[usenames,dvips]{color}
\usepackage{amsmath, mathbbol}

\makeatletter

\@addtoreset{equation}{Section}
\makeatother

\def\lsim{\;\raise0.3ex\hbox{$<$\kern-0.75em\raise-1.1ex\hbox{$\sim$}}\;}
\def\gsim{\;\raise0.3ex\hbox{$>$\kern-0.75em\raise-1.1ex\hbox{$\sim$}}\;}

\def\ben{\begin{enumerate}}  \def\een{\end{enumerate}}
\def\bit{\begin{itemize}}    \def\eit{\end{itemize}}
\def\beq{\begin{equation}}   \def\eeq{\end{equation}}
\def\ba{\begin{array}}       \def\ea{\end{array}}
\def\bea{\begin{eqnarray}}   \def\eea{\end{eqnarray}}

\def\nl{\newline}
\def\ET{$E_\mathrm{T}^\mathrm{miss}$}

\begin{document}

\begin{titlepage}
\renewcommand{\thefootnote}{\fnsymbol{footnote}}
\setcounter{footnote}{0}

\vspace*{-2cm}
\begin{flushright}
LPT Orsay 10-79 
\end{flushright}

\begin{center}
\vspace{2cm}
{\Large\bf Discovering the constrained NMSSM
with tau leptons at the LHC} \\
\vspace{1cm}
{\bf Ulrich Ellwanger$^1$, Alice Florent$^2$, Dirk Zerwas$^2$}\\ 
 \vspace*{.5cm} 
$^1$ Laboratoire de Physique Th\'eorique, UMR 8627, CNRS and
Universit\'e Paris-Sud 11, \\
B\^at. 210, F-91405 Orsay, France\\
$^2$ LAL, Universit\'e Paris-Sud 11, CNRS/IN2P3, Orsay, France
\vspace{2cm}
\end{center}

\begin{abstract}

The constrained Next-to-Minimal Supersymmetric Standard Model (cNMSSM)
with mSugra-like boundary conditions at the GUT scale implies a
singlino-like LSP with a mass just a few GeV below a stau NLSP. Hence,
most of the squark/gluino decay cascades contain two $\tau$ leptons. The
gluino mass $\gsim 1.2$~TeV is somewhat larger than the squark masses of
$\gsim 1$~TeV. We simulate signal and background events for such a
scenario at the LHC, and propose cuts on the transverse momenta of two
jets, the missing transverse energy and the transverse momentum of a
hadronically decaying $\tau$ lepton. This dedicated analysis allows to
improve on the results of generic supersymmetry searches for a large
part of the parameter space of the cNMSSM. The distribution of the
effective mass and the signal rate provide sensitivity to distinguish
the cNMSSM from the  constrained Minimal Supersymmetric Standard Model
in the stau-coannihilation region.

\end{abstract}

\end{titlepage}

\renewcommand{\thefootnote}{\arabic{footnote}}
\setcounter{footnote}{0}

\vspace*{5mm}
\section{Introduction} 

The Next-to-Minimal Supersymmetric Standard Model (NMSSM, for recent
reviews see \cite{Maniatis:2009re,Ellwanger:2009dp}) provides an elegant
solution to the $\mu$-problem of the Minimal Supersymmetric Standard
Model (MSSM) \cite{Kim:1983dt}: after the replacement of the $\mu$-term
in the superpotential of the MSSM by the coupling to a gauge singlet
superfield $S$, the superpotential is scale invariant and the only
dimensionful parameters in the Lagrangian are soft Supersymmetry (SUSY)
breaking terms.

After electroweak symmetry breaking, all components of the gauge
singlet superfield $S$ mix with the components of the MSSM-like Higgs
doublet superfields $H_u$ and $H_d$. Accordingly NMSSM specific
phenomena can take place in the CP-even Higgs sector, the CP-odd Higgs
sector and the neutralino sector. In this paper we concentrate on the
neutralino sector, where the 5th singlet-like neutralino (in addition to
the two neutral higgsinos, the neutral wino and the bino) can be the
Lightest Supersymmetric Particle (LSP)
\cite{Abel:1992ts,Ellwanger:1997jj,Ellwanger:1998vi,
Hesselbach:2000qw,Barger:2006kt} and, simultanously, give rise to a dark
matter density in agreement with WMAP constraints
\cite{Belanger:2005kh,Gunion:2005rw,Cerdeno:2007sn,Hugonie:2007vd,
Belanger:2008nt,Komatsu:2008hk}.

Such a scenario is not far fetched: A conceptually simple origin of soft
SUSY breaking terms is a minimal coupling to supergravity (with a
flavour independent K\"ahler potential and minimal gauge kinetic terms),
and the assumption of spontaneously broken SUSY in a hidden sector.
Then, the soft SUSY breaking terms are universal at the Planck scale
(not far from the GUT scale) in the form of universal scalar masses
$m_0$, universal gaugino masses $M_{1/2}$ and universal trilinear scalar
couplings proportional to $A_0$.

The correspondingly constrained NMSSM (cNMSSM) has been studied first in
\cite{Ellis:1988er,Drees:1988fc,Ellwanger:1993xa,Elliott:1994ht,
King:1995vk,Ellwanger:1996gw}. Since then, the precision of the
radiative corrections has been considerably improved, and the
computation of the dark matter relic density has become possible
\cite{Belanger:2005kh}. Imposing the requirement of a dark matter relic
density in agreement with WMAP constraints \cite{Komatsu:2008hk} as well
as present constraints on Higgs and supersymmetric particle (sparticle)
masses, the parameter space of the cNMSSM has been analysed recently in
\cite{Djouadi:2008yj,Djouadi:2008uj}. (For studies of the
semi-constrained NMSSM, where the singlet-dependent SUSY breaking
parameters are allowed to be non-universal, see
\cite{Ellwanger:2006rn,Djouadi:2008uw,Hugonie:2007vd,Belanger:2008nt,
Balazs:2008ph}.)

Within the cNMSSM, the dark matter constraints require a singlino-like
LSP. The origin of this phenomenon is quite easy to understand: first,
the CP-even scalar singlet $s$ has to assume a non-vanishing vacuum
expectation value (vev) in order to generate the required $\mu$-term.
For this reason its SUSY breaking mass squared $m_S^2$ at the
electroweak scale must not be very large; otherwise the minimum of its
potential is at $s =0$. Second, $m_S^2$ is hardly renormalized between
the GUT and the electroweak scales, which leads to $m_S^2 \sim m_0^2$
with the consequence that, in the cNMSSM, $m_0$ must be small compared
to $M_{1/2}$ and $A_0$.

It is well known that, within the cMSSM \cite{Chung:2003fi}, a small
value of $m_0$ would imply a stau ($\widetilde{\tau}_1$) LSP, which is
not a reasonable candidate for the dark matter. In the cNMSSM, the
singlino-like neutralino $\chi_1^0$ can -- and must -- be lighter than
the $\widetilde{\tau}_1$ for this reason. In order for its dark matter
relic density not being too large, its mass must be only a few GeV below
the mass of the $\widetilde{\tau}_1$ such that
$\chi_1^0-\widetilde{\tau}_1$ coannihilation processes are sufficiently
fast \cite{Hugonie:2007vd,Djouadi:2008yj,Djouadi:2008uj}.

Such a scenario would necessarily have a strong impact on sparticle
searches at colliders (see~\cite{Morrissey:2009tf,Nath:2010zj} for reviews of searches at the LHC): 
since the $\widetilde{\tau}_1$ is the NLSP and 
the singlino-like neutralino $\chi_1^0$
couples only very weakly to all sparticles, the sparticles decay first
into the $\widetilde{\tau}_1$ under the emission of at least
one $\tau$-lepton. 
Subsequently the $\widetilde{\tau}_1$ decays
into $\chi_1^0 + \tau$, hence each sparticle decay cascade contains
typically two $\tau$-leptons. In the present paper we study, for the
first time, the corresponding implications for sparticle searches at the
LHC.

As discussed in \cite{Djouadi:2008yj,Djouadi:2008uj}, the sparticle (and
Higgs) spectrum is quite constrained in the cNMSSM, and can essentially
be parametrized by $M_{1/2}$.
The present lower bounds on $m_{\widetilde{\tau}_1}$ and the lower LEP
bound of $\sim 114$~GeV on the CP-even Standard Model like Higgs mass
require $M_{1/2} \gsim 520$~GeV which, inspite of $m_0$ $\lsim 50$~GeV,
implies a quite heavy sparticle spectrum: squark masses $\gsim 1$~TeV
(apart from the somewhat lighter stop and sbottom masses), and a gluino
mass $\gsim 1.2$~TeV. Since consequently squark production will be
relatively frequent compared to gluino production (in fact, squark +
gluino production dominates), the signal events will mostly lead to two
jets (not counting ISR and FSR) 
with a quite large $p_\mathrm{T}$ as well as a large missing
transverse energy \ET. The sum of the squark and/or gluino production
cross sections is, however, just $\sim 1$~pb or less (for larger
$M_{1/2}$).

Each sparticle production event in the cNMSSM will contain
typically four $\tau$-leptons in the final state: 
two of the $\tau$-leptons (those originating from
$\widetilde{\tau}_1$ decays into $\chi_1^0 + \tau$) will be quite soft
due to the small $\widetilde{\tau}_1 - \chi_1^0$ mass difference $\lsim
5$~GeV and are difficult to detect, whereas the $\tau$-leptons from decays as 
$\chi_2^0 \to \widetilde{\tau}_1 + \tau$ are relatively energetic.

The aim of the present paper is to show that a dedicated analysis 
allows for much better
signal to background ratios for the cNMSSM than standard (generic) supersymmetric 
analyses. We
simulate and study signals and various Standard Model backgrounds for
the LHC at 14~TeV c.m. energy, and find that the signal to background
ratio is sufficiently large allowing for the discovery of the cNMSSM for
a wide range of values of $M_{1/2}$.

Events with \ET, jets and $\tau$-leptons could also be a signal for the
(c)MSSM in the so-called stau-coannihilation region \cite{Ellis:1998kh,
AguilarSaavedra:2005pw, Arnowitt:2006jq, Chattopadhyay:2007di,
Arnowitt:2008bz}. Since the squark/neutralino spectrum of the cMSSM is
necessarily different from the cNMSSM, the combination of
$M_{eff}$ (essentially the sum of the transverse momenta and \ET) and
the signal rate has sensitivity to distinguish the two models.

In the next section we present the spectrum of the cNMSSM. In section~3
we discuss the signal, backgrounds and appropriate cuts. Details of the
simulation of the cNMSSM signal and the Standard Model backgrounds, the
effect of cuts, and the resulting cross sections and signal to
background ratios are given in section~4. Section~5 is devoted to the
comparison of the cNMSSM with the cMSSM, and Section~6 to conclusions
and an outlook.

\section{The spectrum of the cNMSSM}

The NMSSM with a scale invariant superpotential $W$
\cite{Maniatis:2009re,Ellwanger:2009dp} differs from the MSSM through
the replacement of the $\mu$ term in $W_{MSSM}$ by the coupling to a
gauge singlet superfield $S$ and a trilinear $S$ self-coupling:
\beq
\label{eq:2.1}
W_{MSSM} = \mu H_u H_d +\ {\dots} \to W_{NMSSM} = \lambda S H_u H_d +
\frac{\kappa}{3} S^3 +\ \dots\; ,
\eeq
where we have omitted the quark/lepton Yukawa couplings. Hence, if the
vev $s$ of $S$ is non-zero (induced by the soft SUSY breaking terms), an
effective $\mu$ term $\mu_{eff} = \lambda s$ of the desired order of
magnitude is generated and solves the $\mu$ problem of the MSSM
\cite{Kim:1983dt}.

Apart from the Yukawa couplings in the superpotential and the gauge
couplings, the Lagrangian of the NMSSM depends on soft SUSY breaking
scalar masses for the Higgs fields $H_u$, $H_d$ and $S$, the squarks and
the sleptons; trilinear couplings among the scalars (proportional to the
couplings in the superpotential); and gaugino masses for the bino
($M_1$), the winos ($M_2$) and the gluino ($M_3$). Assuming
supersymmetry breaking from a hidden sector in minimal supergravity
(mSUGRA), the SUSY breaking terms are assumed to be universal at the
scale of grand unification (near the Planck scale) and denoted as $m_0$,
$A_0$ and $M_{1/2}$, respectively. Hence the parameters of the
corresponding cNMSSM are, apart from the gauge and quark/lepton Yukawa
couplings,
\beq
\label{eq:2.2}
m_0\; ,\ A_0\; ,\ M_{1/2}\; ,\ \lambda\ \mathrm{and}\ \kappa\; .
\eeq
It is convenient to fix $\kappa$ from the requirement that the Higgs
vevs $h_u$ and $h_d$ generate the correct value of $M_Z$.

As mentioned in the introduction and discussed in detail in
\cite{Djouadi:2008yj,Djouadi:2008uj}, the remaining parameters in
(\ref{eq:2.2}) are strongly constrained: $m_0$ must be small such that
the vev $s$ is non-zero. A small non-zero value for $m_0$ affects
essentially only the singlet-like CP-even Higgs mass
\cite{Djouadi:2008uj}, which is irrelevant for the present study; hence
we assume $m_0 = 0$ in the following. In order to avoid a 
${\widetilde{\tau}_1}$ LSP, the singlino (the fermionic component of
$S$) must be lighter such that ${\widetilde{\tau}_1}$ is the NLSP. The
singlino relic density can be reduced to an amount compatible with WMAP
constraints, if its co-annihilation rate with the ${\widetilde{\tau}_1}$
is large enough, i.e. if the corresponding mass difference is
sufficiently small. This fixes $A_0 \sim -\frac{1}{4} M_{1/2}$ 
\cite{Djouadi:2008yj,Djouadi:2008uj}. Finally $\lambda$ must also be
quite small, since $\lambda$ induces mixings in the CP-even Higgs sector
between the doublet- and singlet-like Higgs states: if the singlet-like
Higgs state is heavier than the (Standard Model like) Higgs state $h$,
the mass of $h$ falls below the LEP bound of $\sim 114$~GeV if $\lambda$
is too large; if the mass of singlet-like Higgs state is below $m_h$
(below $114$~GeV as for the point P520 in Table~\ref{tab:1} below), its
coupling to the $Z$ boson violates again LEP bounds \cite{Schael:2006cr}
for $\lambda$ too large. All in all one finds $\lambda \lsim 0.02$
\cite{Djouadi:2008yj,Djouadi:2008uj} (but $\lambda \gsim 10^{-5}$ in
order still to allow for singlino-${\widetilde{\tau}_1}$
co-annihilation).

$\lambda$ induces also mixings between the singlet-like neutralino and
the MSSM-like neutralinos (bino, neutral wino and higgsinos). For
$\lambda \lsim 0.02$, these mixings are very small. Hence the couplings
of the singlino-like LSP $\chi_1^0$ to all MSSM-like sparticles
(squarks, gluino, sleptons, charginos and neutralinos), which are
induced by these mixings, are very small as well. Accordingly branching
ratios of all these sparticles into the singlino-like LSP are negligibly
small, unless a decay into $\chi_1^0$ is the only decay possible. Due to
R-parity conservation this is the case for the NLSP, the
${\widetilde{\tau}_1}$. Hence sparticle decay cascades proceed as in the
MSSM (with a spectrum as in the cNMSSM, but without the singlet-like
states) until the ${\widetilde{\tau}_1}$ NLSP is produced. Depending on
$\lambda$, the width of the final decay ${\widetilde{\tau}_1} \to
\chi_1^0 + \tau$ can be so small, that the ${\widetilde{\tau}_1}$ decay
vertex is visibly displaced \cite{Djouadi:2008uj}. This case (where the
displaced vertex corresponds to the production of a soft $\tau$-lepton)
could be another interesting signature for the cNMSSM, but subsequently
we will not assume that $\lambda$ is so small that this phenomenon
occurs.

Concerning the remaining parameter $M_{1/2}$, we find that the LEP
constraints on the Higgs sector require $M_{1/2} \gsim 520$~GeV. Then,
all bounds on sparticle masses from colliders as well as constraints
from B-physics are satisfied. For the calculation of the
spectrum we use the code NMSPEC \cite{Ellwanger:2006rn} within
NMSSMTools \cite{Ellwanger:2004xm,Ellwanger:2005dv}, updated including
radiative corrections to the Higgs sector from \cite{Degrassi:2009yq}.
The dark matter relic density is computed with the help of micrOMEGAs
\cite{Belanger:2005kh}. Clearly, very large values of $M_{1/2}$ are
generally disfavoured by fine-tuning arguments; moreover, smaller values
of $M_{1/2}\lsim 1$~TeV allow to explain the
discrepancy of the measured anomalous magnetic moment of the muon with
the Standard Model \cite{Djouadi:2008uj}. Subsequently we confine
ourselves to $M_{1/2} \leq 1$~TeV. In Table~\ref{tab:1} we show the values
for $A_0$, $\tan\beta = h_u/h_d$ and $\mu_{eff}$ ($A_0$ is determined by
the correct dark matter relic density, whereas $\tan\beta$ and
$\mu_{eff}$ are obtained as output) as well as the Higgs and sparticle
spectra for $M_{1/2} = 520,\ 600,\ 800$~GeV and 1~TeV for $m_0 = 0$ and
$\lambda= 0.001$.

\begin{table}[!ht]
\begin{center}
\begin{tabular}{|l|c|c|c|c|}
\hline
&\phantom{eV}P520\phantom{eV}
&\phantom{eV}P600\phantom{eV}
&\phantom{eV}P800\phantom{eV}
&\phantom{eV}P1000\phantom{eV}
\\
\hline\hline
$M_{1/2}$ (GeV) & 520  &    600 & 800 & 1000 \\
\hline
$A_0$ \ \ \ (GeV) &-142  & -166 & -225 & -282\\
\hline
$\tan \beta$ & 23.2 &  24.3 & 26.6 & 28.3\\
\hline
$\mu_\mathrm{eff}$ \ \ \  (GeV)& 666 & 757 & 977 & 1190 \\
\hline\hline
$m_{h_1^0}$ \ \ \ (GeV) &  100 &  115 & 117 & 118\\
\hline
$m_{h_2^0}$\ \ \  (GeV) &  115&    118 & 159 & 199 \\
\hline
$m_{h_3^0}$ \ \ \ (GeV) &  654&    738 & 937 & 1127 \\
\hline
$m_{a_1^0}$\ \ \  (GeV) &  174 & 203 & 275 & 345 \\
\hline
$m_{a_2^0}$\ \ \  (GeV) &  654 & 738 & 937 & 1127 \\
\hline
$m_{h^\pm}$\ \ \  (GeV) &  667 & 751 & 951 & 1140 \\
\hline\hline
$m_{\chi_1^0}$\ \ \  (GeV) &  142 & 166 & 225 & 282 \\
\hline
$m_{\chi_2^0}$ \ \ \ (GeV) &  215 & 250 & 338 & 427 \\
\hline
$m_{\chi_3^0}$\ \ \  (GeV) &  404 & 471 & 636 & 801 \\
\hline
$m_{\chi_{4,5}^0}$\ \ \  (GeV)  & 680 &  770 & 990 & 1200 \\
\hline
$m_{\chi^\pm_1}$\ \ \  (GeV)  & 404 & 471 & 636 & 801 \\
\hline
$m_{\chi^\pm_2}$\ \ \  (GeV)  &  684 &  773 & 992 & 1203 \\
\hline\hline
$m_{\tilde g}$ \ \ \ (GeV)  & 1192 & 1361 & 1777 & 2187 \\
\hline
$m_{\tilde u_L}$\ \ \  (GeV)  & 1082 & 1234 & 1607 & 1973 \\
\hline
$m_{\tilde u_R}$\ \ \  (GeV)  & 1044 & 1189 & 1546 & 1895 \\
\hline
$m_{\tilde d_L}$\ \ \  (GeV)  & 1085 & 1237 & 1609 & 1974 \\
\hline
$m_{\tilde d_R}$\ \ \  (GeV)  & 1040 & 1184 & 1539 & 1886 \\
\hline
$m_{\tilde t_1}$ \ \ \ (GeV)  & 825 & 947 & 1246 & 1538 \\
\hline
$m_{\tilde t_2}$\ \ \  (GeV)  & 1032 & 1165 & 1492 & 1816 \\
\hline
$m_{\tilde b_1}$\ \ \  (GeV)  & 973 & 1109 & 1444 & 1772 \\
\hline
$m_{\tilde b_2}$\ \ \  (GeV)  & 1020 & 1158 & 1496 & 1826 \\
\hline\hline
$m_{\tilde e_L}$\ \ \  (GeV)  & 347 & 399 & 527 & 654 \\
\hline
$m_{\tilde e_R}$\ \ \  (GeV)  & 196 & 224 & 296 & 368 \\
\hline
$m_{\tilde \nu_l}$\ \ \  (GeV)  & 338 & 391 & 521 & 650 \\
\hline
$m_{\tilde \tau_1}$\ \ \  (GeV)  & 147 & 171 & 229 & 286 \\
\hline
$m_{\tilde \tau_2}$\ \ \  (GeV)  & 353 & 403 & 525 & 647 \\
\hline
$m_{\tilde \nu_\tau}$\ \ \  (GeV)  & 332 & 383 & 509 & 633 \\
\hline
\hline
$\sigma$ (pb) & 1.36  &  0.70 & 0.134 & 0.035 \\
\hline
\end{tabular}
\end{center}
\caption{Input parameters, $\tan\beta$, $\mu_{eff}$ and low-energy
spectra for four points of the \nobreak{cNMSSM} with $m_0 = 0$ and
$\lambda = 0.001$. In the last line we give the total NLO cross sections
for the production of all sparticles at the LHC.}
\label{tab:1} 
\end{table}

We see that, as announced above, $m_{\widetilde \tau_1}- m_{\chi_1^0}
\lsim 5$~GeV and hence $\tau$-leptons from the decay ${\widetilde
\tau_1} \to {\chi_1^0} + \tau$ are necessarily soft. $\tau$-leptons from
the decay $\chi_2^0 \to \widetilde \tau_1 + \tau$ (where $\chi_2^0$ is
dominantly bino-like) profit at least from $m_{\chi_2^0} - m_{\widetilde
\tau_1} \sim 70$~GeV (for P520), or more energy from the decays of other
sparticles into ${\widetilde \tau_1}$. Note that right-handed sleptons
$\widetilde e_R$ and $\widetilde \mu_R$ decay essentially via the
three-body channel as $\widetilde e_R \to e +\widetilde\tau_1+ \tau$. In
fact, apart from the NMSSM-specific decay ${\widetilde \tau_1} \to
{\chi_1^0} + \tau$ (with a branching ratio of 100~\%), the corresponding
sparticle decay branching ratios can be obtained from the code SUSY-HIT
\cite{Djouadi:2006bz} and the MSSM with a corresponding spectrum, which
is used for the simulations of events below.

At the LHC, the dominant sparticle production processes are of course
squark-gluino and squark-(anti-)squark pair productions. Subsequently a
typical squark decay cascade looks like
\beq
\label{eq:2.3}
\widetilde{q} \to q + \chi_2^0 \to q + \tau+\widetilde{\tau_1} \to q +
\tau+\tau+\chi_1^0\; ,
\eeq
but many more possibilities exist. Their simulation, together with the
simulation of Standard Model background processes, will be discussed in
the next sections.

\section{The signal, backgrounds and cuts}

As for most SUSY models, the production of squarks of the first
generation and of gluinos will be the dominant sparticle production
processes at the LHC. Their total production cross sections are
obtained at NLO (QCD) from PROSPINO \cite{Beenakker:1996ed}, and are
also shown in Table~\ref{tab:1}. The
dominant contributions originate from squark + gluino production ($\sim
37\%$) and squark pair production ($\sim 25\%$); less dominant are
squark + antisquark production ($\sim 13\%$) and gluino pair production
($\sim 5\%$). The production cross sections of stop and sbottom squarks,
sleptons, charginos and neutralinos add another $\sim 19\%$ to the total
sparticle production cross sections.

The dominant background processes for SUSY searches are well-known:
top-antitop pair production, W+n-jet production, Z+n-jet
production, W+Z production and WW+n-jet production. Since we will
compare the performance of our simulation with the results of standard
SUSY searches by ATLAS~\cite{Aad:2009wy}, we assume the same production
cross sections for these background processes as in~\cite{Aad:2009wy}
(given in Table~\ref{tab:AcerDetbg} below).

Given that
gluinos (whose decay generates typically two hard jets) are even
somewhat heavier than the first generation squarks (generating typically
one hard jet), we require at least two hard jets per event only whereas in generic
supersymmetric analyses usually four hard jets are required. On
the other hand, given the large squark and gluino masses, we can require
quite large transverse momenta of the jets as well as a large missing
transverse energy $E_\mathrm{T}^\mathrm{miss}$.

For the $\tau$-leptons we consider their hadronic decays only. For their
transverse momenta we require at least 30~GeV, which allows to assume an
efficiency of $\sim 40\%$~\cite{Aad:2009wy} (and a $\tau$-fake rate of
jets of $\sim 1-2\%$). On the other hand, since only two among the four
$\tau$-leptons per event are sufficiently energetic and the total signal
cross sections are already quite small, we require one identified
$\tau$-lepton only.

Additional standard cuts are a lower limit on the angle $\Delta \Phi$
between the hard jets and $E_\mathrm{T}^\mathrm{miss}$, as well as a cut
on the transverse mass $M_T$ formed from $E_\mathrm{T}^\mathrm{miss}$
and the identified $\tau$-lepton (in order remove semileptonically decaying 
W$+$jets events). Altogether, the list of our cuts is given by:

\begin{enumerate}
\item At least two jets, one with $p_\mathrm{T} > 300$~GeV and one with $p_\mathrm{T} >
150$~GeV

\item $E_\mathrm{T}^\mathrm{miss} > 300$~GeV

\item At least one $\tau$-lepton with $p_\mathrm{T} > 30$~GeV

\item $\Delta \Phi(j_i,E_\mathrm{T}^\mathrm{miss}) > 0.2$ for
the hard jets

\item $M_T > 100$~GeV, where $M_T$ is computed from the visible momenta
of the hardest $\tau$-lepton and $E_\mathrm{T}^\mathrm{miss}$.
\end{enumerate}

Below we will denote this set of cuts as cNMSSM analysis.

\section{Simulation, signal and background rates}

Both the signal and the top quark background were generated by
PYTHIA~6.4~\cite{Sjostrand:2006za}, which was in charge of
generation and phase-space decays. PYTHIA performed the parton
showering as well as the matching procedure according to the MLM
prescription including initial and final state radiation. 
TAUOLA~\cite{Jadach:1990mz,Jezabek:1991qp,Jadach:1993hs} was employed
for the $\tau$-decays. All other backgrounds (involving at least one W
or Z-boson) were generated with ALPGEN~\cite{Mangano:2002ea}. In order
to keep the statistics manageable, preselection cuts were applied on the
ALPGEN samples. Since event generation was performed by leading order
generators, the cross sections were scaled according to the NLO cross
sections in Table~\ref{tab:1} for the signal, as in~\cite{Aad:2009wy}
to the NLO$+$NLL calculation for top production, and to NLO (or NNLO
level, where available) for electroweak boson(s) production.

For the detector simulation we employed
AcerDet~\cite{RichterWas:2002ch}. AcerDet is a fast detector simulation
which provides a reasonable  description of the performance of an LHC
detector. The events generated by PYTHIA and ALPGEN+PYTHIA were all
passed through AcerDet. AcerDet reconstructs jets, leptons and the
missing transverse energy, it also labels the origin of the jets, e.g.,
those coming from a tau lepton. The efficiency and the corresponding
background rejection for a working point of 40\% $\tau$ identification
efficiency were implemented at reconstruction level.

One of the issues to be checked is the energy of the reconstructed
tau-jets. For this initial check we used PYTHIA to produce Z bosons and
their subsequent decay to tau leptons. The hadron-hadron as well  as the
lepton-hadron final states were reconstructed and compared to the
results of the ATLAS collaboration presented in~\cite{Aad:2009wy}.
Reasonable agreement at the level of a few percent (about 3~GeV) was
found. 

A fast detector simulation as AcerDet will not be able to simulate the
non-Gaussian tails, e.g., in the transverse missing energy
distribution. Simulation and reconstruction results without full
detector simulation and reconstruction are therefore not expected to be
perfectly reproduced. To get a feeling of how well the background can be
modeled with AcerDet, two signatures of Ref.~\cite{Aad:2009wy} have been
implemented and analysed in addition to the dedicated cNMSSM analysis:
the four-jet SUSY search (4j0l) and the SUSY search with at least one
tau in the final state (4jtau). 

\begin{table}[hb!]
\begin{tabular}{|c|cc|c|c|c|}
\hline
               &        Events     &    cross section (pb)  &    4j0l  (fb) &   4jtau  (fb)    &     cNMSSM (fb)   \\
\hline
tt             &        2110000    &            833         &       350$\pm$12      &         50$\pm$4.4       &         7.9$\pm$1.7         \\
\hline
W+2jets        &         222700    &            281         &              0        &                0         &                0          \\
W+3jets        &          89250    &            116         &         15$\pm$4.5    &                0         &         2.6$\pm$1.8         \\
W+4jets(inc)   &          50875    &             61         &        220$\pm$16     &         9.6$\pm$3.4      &         4.8$\pm$2.4           \\
\hline
W+jets (inc)   &                   &                        &        235$\pm$16.6   &         9.6$\pm$3.4       &        7.4$\pm$3        \\
\hline
Z+2jets        &          88850    &             106        &          0            &                0         &                0          \\
Z+3jets        &          22320    &             27.5       &          0            &                 0        &                 0         \\
Z+4jets(inc)   &          11639    &             10.1       &      0.9$\pm$0.9      &                 0        &                 0         \\
\hline
Z+jets         &                   &                        &      0.9$\pm$0.9      &                 0        &                 0         \\
\hline
ZW(inc)       &            250    &               0.5       &           2$\pm$2     &                  0       &                  0        \\
\hline
WW+0jet        &          50000    &             47         &           0           &                  0       &                  0        \\
WW+1jet        &          15000    &             20         &           0           &                  0       &                  0        \\
WW+2jet(inc)   &          32796    &             13         &      17.8$\pm$2.6     &             1.2$\pm$0.7 &          0.8$\pm$0.6  \\
\hline
WW+jets (inc)  &                   &                        &      17.8$\pm$2.6     &           1.2$\pm$0.7    &          0.8$\pm$0.6     \\
\hline\hline
Total          &                   &                        &      606$\pm$21       &             61$\pm$5.6    &         16.1$\pm$3.5   \\
\hline\hline   
\end{tabular}
\caption{\label{tab:AcerDetbg} The number of simulated background
events, cross sections before cuts, and cross sections after the 4j0l,
4jtau and the cNMSSM analysis. The quoted error is the statistical error.}
\end{table}

In Table~\ref{tab:AcerDetbg} the result of the background cross sections
before and after the cuts is shown for the two ATLAS as well as the
cNMSSM signatures. For the 4j0l analyses, the AcerDet-result for the
total background cross section of about 606~fb after cuts is of the
right order of magnitude compared to 708~fb (NLO) in~\cite{Aad:2009wy}.
Adding the tau identification to the analysis in the 4jtau analysis, a
total background cross section of 61~fb after cuts obtained here is to
be compared with 51~fb at NLO in~\cite{Aad:2009wy}. Thus with and
without tau identification, AcerDet provides a reasonable estimate of the
expected background with respect to the dedicated full simulation.

We note that already with the softer cuts within the 4j0l analyses and
notably the cuts on $\Delta \Phi(j_i,E_\mathrm{T}^\mathrm{miss})$, the
remaining QCD background was found to be small in~\cite{Aad:2009wy}.
In our case, QCD events could pass the cNMSSM cuts only if a very large value
of $E_\mathrm{T}^\mathrm{miss}$ {\it and} a $\tau$-lepton would be faked
simultaneously. Assuming a jet $\to$ $\tau$ fake rate up to $\sim 2\%$
(for an acceptance of 40\%), and that the suppression rate of QCD events
without missing energy for $E_\mathrm{T}^\mathrm{miss})> 150$~GeV is 1\%~\cite{Aad:2009wy}
while we cut at 300~GeV, we should be safe of the QCD background.

Whereas the two ATLAS analyses are designed to cover a large variety of
supersymmetric signals, the signatures discussed above are chosen
specifically for heavy squarks and gluinos as well as $\tau$-rich final
states as in the cNMSSM. The background cross sections for this analysis
are shown in the last column of Table~\ref{tab:AcerDetbg}. The total
background, already decreased from the 4-jet-0-lepton to the 4-jet-tau
analysis by an order of magnitude, is reduced by another factor four to
$\sim 16$~fb. 

Typically the overall efficiency for all SUSY processes weighted by the
cross section varies between 7\% and 10\%.  The cross section for the
cNMSSM benchmark points after all cuts are shown in
Table~\ref{tab:signal} for the three analyses. The S/B ratio, the ratio 
S/$\sqrt{\mathrm{B}}$ for an integrated luminosity of 1~fb$^{-1}$  as well as
for 30~fb$^{-1}$ are shown. 

\begin{table}[h!]
\begin{center}
\begin{tabular}{|c|c|c|c|c|}
\hline
        &               & 4j0l       & 4jtau     & cNMSSM \\    
\hline
P520         &  1.36pb      &      101$\pm$1.1fb    &  27$\pm$0.5fb     &    99$\pm$1fb     \\
S/B          &              &        0.17           &      0.44         &     6.2             \\
S/$\sqrt{\mathrm{B}}$ & 1~fb$^{-1}$  &        4.1            &      3.4          &     24.8             \\
S/$\sqrt{\mathrm{B}}$ & 30~fb$^{-1}$ &        22             &      19           &     136             \\
\hline
P600         &   0.70pb     &      47$\pm$0.5fb    &  13.3$\pm$0.3fb   &    58$\pm$0.6fb   \\
S/B          &              &        0.08          &    0.21           &       3.6           \\
S/$\sqrt{\mathrm{B}}$ & 1~fb$^{-1}$  &        1.9           &   1.7             &       14.5             \\
S/$\sqrt{\mathrm{B}}$ & 30~fb$^{-1}$ &        10.5          &   9.3             &       79             \\
\hline
P800         &   0.134pb    &       7.5$\pm$0.1fb  &  2.7$\pm$0.05fb   &   13.4$\pm$0.1fb   \\
S/B          &              &         0.012        &   0.04            &    0.8             \\
S/$\sqrt{\mathrm{B}}$ & 1~fb$^{-1}$  &         0.3          &     0.34          &    3.4               \\
S/$\sqrt{\mathrm{B}}$ & 30~fb$^{-1}$ &         1.7          &     1.9           &     18               \\
\hline
P1000        &    0.035pb   &       1.68$\pm$0.02fb   &  0.62$\pm$0.01fb   &  3.43$\pm$0.03fb  \\
S/B          &              &          0.002          &       0.01        &    0.2             \\
S/$\sqrt{\mathrm{B}}$ & 1~fb$^{-1}$  &          0.07           &       0.07         &    0.86               \\
S/$\sqrt{\mathrm{B}}$ & 30~fb$^{-1}$ &          0.37           &       0.43         &    4.7              \\
\hline
\end{tabular}
\caption{\label{tab:signal} Signal expectation for the NMSSM points at
NLO after all cuts for the benchmark points. At least 120000 events per
point were generated. The error is the statistical error. For every
point the ratios S/B and S/$\sqrt{\mathrm{B}}$ for an integrated
luminosity of 1~fb$^{-1}$ and 30~fb$^{-1}$ are shown.}
\vspace*{-2mm}
\end{center}
\end{table}


Table~\ref{tab:signal} clarifies the advantage of the cNMSSM cuts with
respect to the general analysis: the ratio S/$\sqrt{\mathrm{B}}$ is
$7-10$ times larger for the cNMSSM cuts even with respect to the
standard 4-jet-tau signature, which originates both from the larger
background suppression and the larger efficiency on the signal.
Correspondingly the cNMSSM cuts allow for a sensitivity, for a given
luminosity, on a much larger part of the cNMSSM parameter space (for
heavier squarks/gluinos). The point P800 (with squark/gluino masses of
1.6/1.7~TeV) is hardly visible within the standard analysis even for
30~fb$^{-1}$, whereas the ratio S/$\sqrt{\mathrm{B}}$ is still $\sim 18$
for the cNMSSM cuts. Only for the point P1000 (with squark/gluino masses
of 1.9/2.2~TeV) a larger luminosity and/or even harder cuts seem to be
required for detection.

\begin{figure}[htb]
\resizebox{1.0\textwidth}{!}{%
  \includegraphics{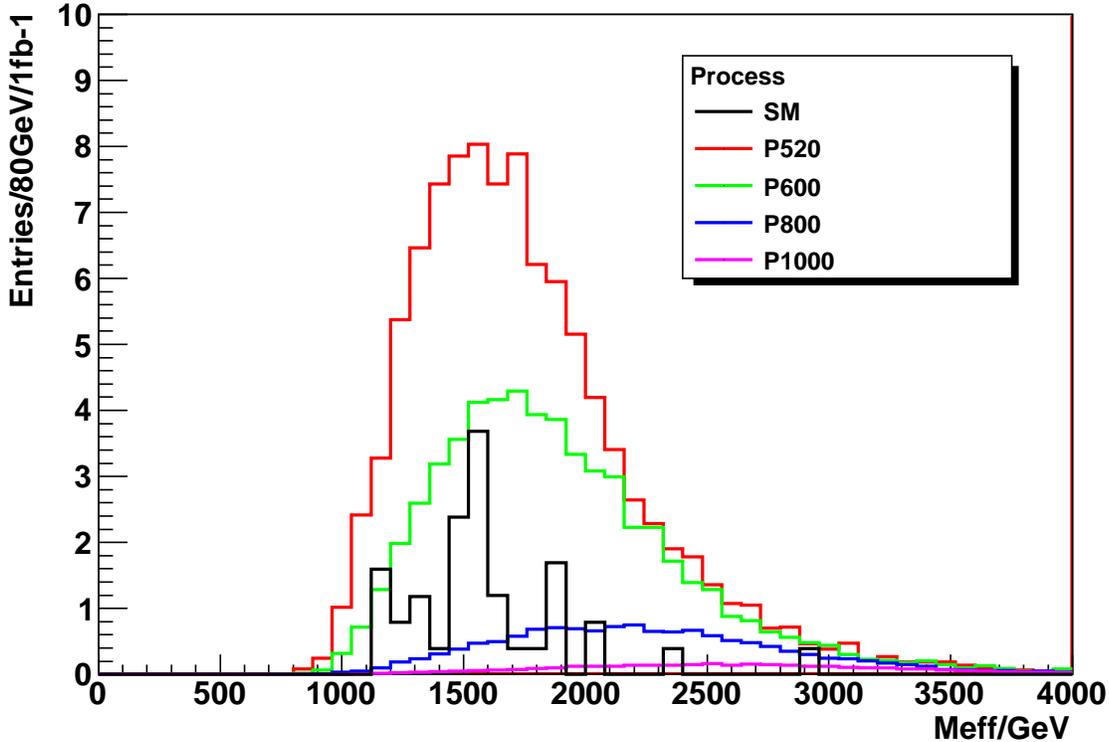}
}
\caption{The effective mass distribution for the SM background and the
cNMSSM points from Table~1, after all cuts of the cNMSSM analysis are
applied, normalised to an integrated luminosity of 1~fb$^{-1}$.}
\label{fig:meff}       
\end{figure}

In Figure~\ref{fig:meff} we show the spectrum of the effective mass
$M_\mathrm{eff} \equiv \sum p_\mathrm{T}^\mathrm{jets} + \sum
p_\mathrm{T}^\mathrm{lep} + E_\mathrm{T}^\mathrm{miss}$ after all cuts
of the cNMSSM analysis are applied, normalised to an integrated
luminosity of 1~fb$^{-1}$. Typically, the spectrum of the effective mass
peaks at a value corresponding to the masses of the pair produced
sparticles~\cite{Aad:2009wy}. Here the maxima of $M_\mathrm{eff}$ are
shifted to somewhat larger values due to the cuts on
$p_\mathrm{T}^\mathrm{jets}$ and $E_\mathrm{T}^\mathrm{miss}$. As expected,
the spectrum of $M_\mathrm{eff}$ is harder for the points with heavier
squarks/gluinos.

\begin{figure}[htb]
\resizebox{1.0\textwidth}{!}{%
  \includegraphics{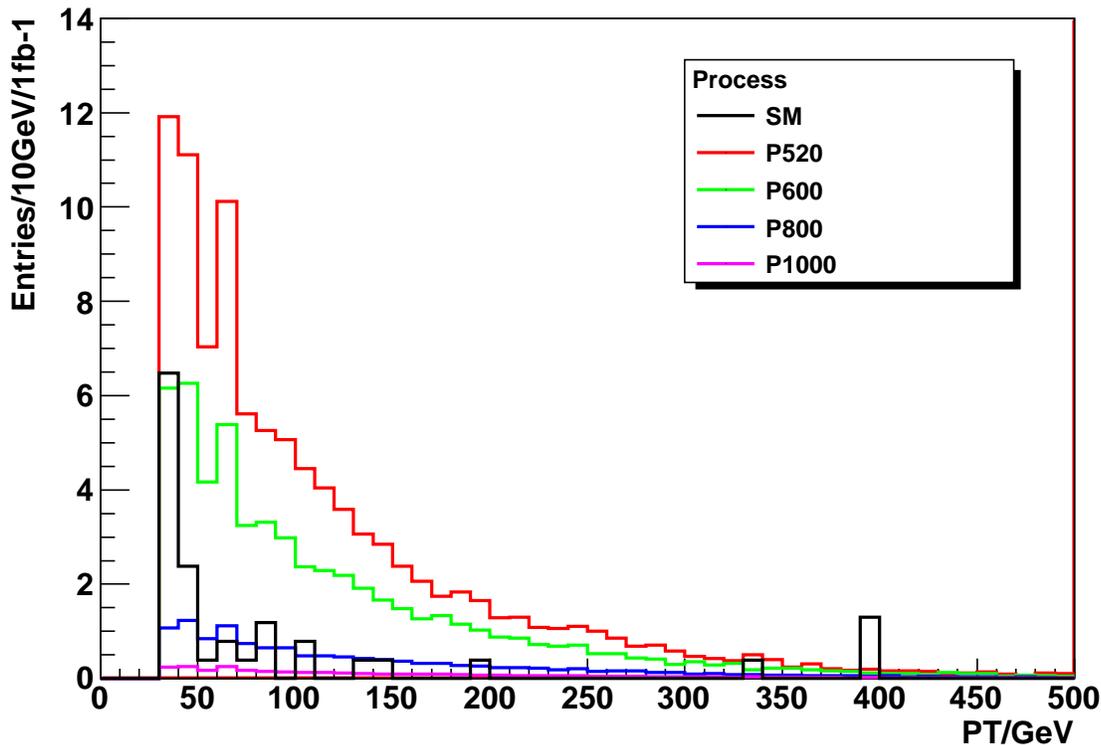}
}
\caption{The transverse momentum distribution of the leading tau
after all cuts of the cNMSSM analysis are applied,
normalised to an integrated luminosity of 1~fb$^{-1}$.}
\label{fig:pt}      
\end{figure}

For completeness we show in Figure~\ref{fig:pt} the transverse momentum
of the leading tau candidate (after the cNMSSM cuts). Modulo the rate,
the spectrum of the leading tau candidate is slightly harder for the
points with heavier squarks/gluinos. 

\section{Comparison with the cMSSM in the stau coannihilation region}

It is well-known that $\tau$-rich final states from squark or gluino
production would also be generated in the so-called stau coannihilation
region of the (c)MSSM \cite{Ellis:1998kh, AguilarSaavedra:2005pw,
Arnowitt:2006jq, Chattopadhyay:2007di, Arnowitt:2008bz}. Hence the
question arises, by which signatures this region of the cMSSM can be
distinguished from the cNMSSM. 

Clearly, the neutralino/$\widetilde{\tau}$ spectrum of the cMSSM is
different from the one of the cNMSSM: in the cMSSM (in the stau
coannihilation region, which we assume henceforth) the lighter
$\widetilde{\tau}_1$ has a mass close to the bino-like neutralino LSP
$\chi_1^0$, whereas the neutralino $\chi_2^0$ is typically wino-like.
At first sight, a squark decay cascade as in Eq.(\ref{eq:2.3}) is
also possible within the cMSSM, with corresponding replacements of the
neutralinos $\chi_1^0$ and $\chi_2^0$. However, all right-handed squarks
(and sleptons) would not couple to the wino-like $\chi_2^0$, and prefer
to decay directly into the bino-like $\chi_1^0$. These decays do {\it
not} lead to two $\tau$-leptons in the cascade. As a consequence, the
$\tau$-rich cascades hardly occur for right-handed squark decays, and
are thus less frequent (relative to the total squark production cross
section) than in the cNMSSM.

On the other hand, squarks and gluinos can be considerably lighter
in the cMSSM than in the cNMSSM, since smaller values of $M_{1/2}$ --
together with larger non-zero values of $m_0$ -- are allowed. In
particular this is the case, if we look for a point in the cMSSM
parameter space with similar $\chi_1^0$ and $\widetilde{\tau}_1$ masses
as the point P520 of the cNMSSM. It turns out that, for the
corresponding values of $M_{1/2}$ and $m_0$ (we take $A_0=0$ for
simplicity), the squarks and gluinos are considerably lighter than
for the point P520. Hence we denote this point as MSSMl (``l'' for
``light''). Its parameters and sparticle masses are given in
Table~\ref{tab:mssmspec} below.

\begin{table}[!h]
\begin{center}
\begin{tabular}{|l|c|c|c|c|}
\hline
&\phantom{eV}MSSMl\phantom{eV}
&\phantom{eV}MSSMh\phantom{eV}
\\
\hline
$M_{1/2}$ (GeV) & 360  &    520  \\
\hline
$m_0$\ \ \ (GeV) &210&200\\
\hline
$A_0$ \ \ \ (GeV) & 0  & 0 \\
\hline
$\tan \beta$ & 40 &  30 \\
\hline
$\mu_\mathrm{eff}$ \ \ \  (GeV)& 466 & 649  \\
\hline\hline
$m_{h_1^0}$ \ \ \ (GeV) &  114 &  117\\
\hline\hline
$m_{\chi_1^0}$\ \ \  (GeV) &  146 & 215 \\
\hline
$m_{\chi_2^0}$ \ \ \ (GeV) &  274 & 406  \\
\hline
$m_{\chi_{3,4}^0}$\ \ \  (GeV) &  480 & 660 \\
\hline
$m_{\chi^\pm_1}$\ \ \  (GeV)  & 274 & 406 \\
\hline
$m_{\chi^\pm_2}$\ \ \  (GeV)  &  486 &  666 \\
\hline\hline
$m_{\tilde g}$ \ \ \ (GeV)  & 851 & 1191\\
\hline
$m_{\tilde u_L}$\ \ \  (GeV)  & 800 & 1100 \\
\hline
$m_{\tilde u_R}$\ \ \  (GeV)  & 776 & 1062  \\
\hline
$m_{\tilde d_L}$\ \ \  (GeV)  & 804 & 1103 \\
\hline
$m_{\tilde d_R}$\ \ \  (GeV)  & 774 & 1058  \\
\hline
$m_{\tilde t_1}$ \ \ \ (GeV)  & 598 & 843 \\
\hline
$m_{\tilde t_2}$\ \ \  (GeV)  & 765 & 1038 \\
\hline
$m_{\tilde b_1}$\ \ \  (GeV)  & 688 & 984 \\
\hline
$m_{\tilde b_2}$\ \ \  (GeV)  & 749 & 1033  \\
\hline\hline
$m_{\tilde e_L}$\ \ \  (GeV)  & 322 & 401 \\
\hline
$m_{\tilde e_R}$\ \ \  (GeV)  & 252 & 280 \\
\hline
$m_{\tilde \nu_l}$\ \ \  (GeV)  & 312 & 393\\
\hline
$m_{\tilde \tau_1}$\ \ \  (GeV)  & 156 & 222\\
\hline
$m_{\tilde \tau_2}$\ \ \  (GeV)  & 332 & 405 \\
\hline
$m_{\tilde \nu_\tau}$\ \ \  (GeV)  & 294 & 382 \\
\hline
\hline
$\sigma$ (pb) & 9.44  &  1.40  \\
\hline
\end{tabular}
\caption{\label{tab:mssmspec} SUSY breaking parameters, $\tan\beta$,
$\mu_\mathrm{eff}$, sparticle spectra and total sparticle cross sections
for the cMSSM points MSSMl and MSSMh.}
\end{center}
\end{table}

As indicated in the last line in Table~\ref{tab:mssmspec}, the lighter
squarks and gluinos imply considerably larger production cross sections
for the point MSSMl compared to P520. As a consequence, the number of
events passing our cNMSSM analysis above is larger than for P520, in spite
of the absence of $\tau$-leptons in right-handed squark decays. 

There exist also points in the cMSSM parameter space where the
squark and gluino spectrum resembles the one of P520, implying similar
production cross sections. Such points correspond to larger values of
$M_{1/2}$ and $m_0$; an example is given by the point MSSMh (``h'' for
``heavy''), whose squark and gluino masses are similar to those of the
cNMSSM point P520 (see Table~\ref{tab:mssmspec}). 

The signal rates after the cNMSSM cuts for the points MSSMl and MSSMh
are given in Table~\ref{tab:mssmsignal}: these are considerably larger
(as compared to P520) for the point MSSMl, but smaller for the point
MSSMh in spite of the similar squark/gluino masses and hence the similar
total sparticle cross section (see Table~\ref{tab:mssmspec}). The reason
was  mentioned above: right-handed squarks do not decay via $\tau$-rich
cascades and, hence, right-handed squark decays do not contribute to the
signal after the cNMSSM analysis.

\begin{table}[htb]
\begin{center}
\begin{tabular}{|c|c|c|c|c|}
\hline
        &               & 4j0l       & 4jtau     & cNMSSM analysis \\    
\hline
MSSMl        &  9.4pb       &     1429$\pm$10fb  &  121$\pm$3fb       &   166$\pm$4fb   \\
S/B          &              &       2.4          &    2.0             &   10             \\
S/$\sqrt{\mathrm{B}}$ & 1~fb$^{-1}$  &       58           &    15              &   42             \\
S/$\sqrt{\mathrm{B}}$ & 30~fb$^{-1}$ &       320          &    84              &   227             \\
\hline
MSSMh        &   1.40pb     &     242$\pm$1.7fb      &  22$\pm$0.5fb     &    41$\pm$0.7fb   \\
S/B          &              &         0.4            &   0.46            &   2.5             \\
S/$\sqrt{\mathrm{B}}$ & 1~fb$^{-1}$  &         9.8            &    2.8            &    10               \\
S/$\sqrt{\mathrm{B}}$ & 30~fb$^{-1}$ &         54             &    15             &    56               \\
\hline
\hline
\end{tabular}
\caption{\label{tab:mssmsignal} Signal expectation for the MSSM points
at NLO after all cuts. 120000 events per point
minimum were generated. The error is the statistical error.}
\vspace*{-5mm}
\end{center}
\end{table}

\begin{figure}[h!]
\begin{center}
\resizebox{1.0\textwidth}{!}{%
  \includegraphics[width=0.7\linewidth]{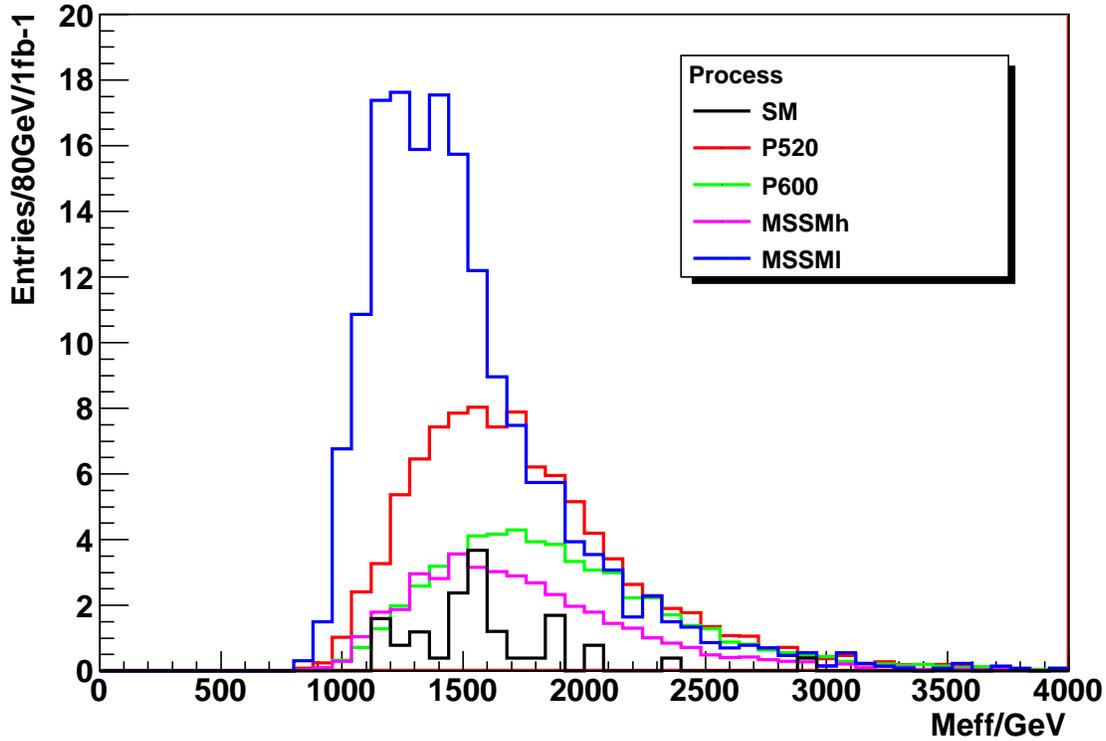}
  }
\end{center}
\caption{The effective mass distribution for the points P520, P600,
MSSMl and MSSMh after all cuts of the cNMSSM analysis are applied.}
\label{fig:Meff_mssm}   
\end{figure}

The $M_\mathrm{eff}$ spectrum for the MSSM points, together with P520,
P600 and the SM background, is shown in Fig.~\ref{fig:Meff_mssm}.
First, the point MSSMl (with similar $\chi_1^0$ and $\widetilde{\tau}_1$
masses as the point P520) has not only a larger signal cross section as
compared to P520, but we see that its maximum of $M_\mathrm{eff}$ is
visibly shifted towards smaller values. 

Can we distinguish the point MSSMh from any of the cNMSSM points? Due to
the similar squark/gluino masses as P520, MSSMh has its maximum of
$M_\mathrm{eff}$ in the same region as P520, but a significantly smaller
signal rate. The signal rate for the cNMSSM point P600 of about 58~fb is
still larger than 41~fb for MSSMh. (The difference is slightly larger 
than a conservative error of 20-30\% on the theoretical cross section
prediction.) On the other hand we see in Fig.~\ref{fig:Meff_mssm} that
the maximum of $M_\mathrm{eff}$ for P600 is shifted towards larger
values due to the heavier squarks/gluinos: the root mean square of the
distributions is about 500~GeV and the difference of the average effective mass is about
130~GeV, so that the error on the average effective mass for 1~fb$^{-1}$ is
about 70~GeV, i.e., about two times smaller than the
difference. The average effective mass is affected somewhat by the tails
for large effective mass. Using a simple fit of the distributions, the peak to peak difference
increases slightly to about 150~GeV providing for a stronger separation 
of MSSMh and P600. 
Any cNMSSM point with still heavier squarks/gluinos (such
that the signal rate coincides with the one for MSSMh) will imply a
maximum for still larger values of $M_\mathrm{eff}$.
Hence, the cNMSSM points have either measurably larger
signal rates after applying the cNMSSM cuts (if the maxima of
$M_\mathrm{eff}$ coincide with a MSSM point), or maxima at measurably
larger values of $M_\mathrm{eff}$ (if the signal rates coincide with a
MSSM point).

Additionally one can compare the cross section ratios after the generic
supersymmetric 4j0l cut, which are 242~fb (MSSMh, where squarks are
somewhat lighter than gluinos) as compared to 47~fb (P600, where gluinos
are somewhat heavier than squarks). Hence, given a corresponding signal
in the data, a careful comparison of both the signal rates for generic
and dedicated searches and the maximum of $M_\mathrm{eff}$ (possibly
including in addition the transverse momentum of the tau lepton)
should allow to distinguish the cNMSSM from the MSSM in the stau
coannihilation region.


\section{Conclusions and outlook}

In the present paper we have proposed criteria for the search for the
fully constrained NMSSM at the LHC. In view of the relatively heavy
squarks and gluinos in the cNMSSM and correspondingly small production
cross sections, this task is not quite trivial. On the other hand, due
to the large number of $\tau$-leptons in the final states, signatures
involving hadronic $\tau$ decays are relatively efficient. Whereas the
soft $\tau$-leptons in the final states are difficult to use, the
requirement of at least one hard $\tau$-lepton has a relatively large
signal acceptance.

Combining this requirement with relatively hard cuts on the transverse
momenta of two jets and $E_\mathrm{T}^\mathrm{miss}$ as specified at the
end of section~3, the signal to background ratio is significantly
improved with respect to the more standard 4j0l or 4jtau analyses. This
result was obtained after simulations including detector effects and a
$\tau$ acceptance, which we compared with and checked against the
analysis of SUSY signals by the ATLAS group. Hence, already an
integrated luminosity of 1~fb$^{-1}$ (at 14~TeV c.m. energy) becomes
sensitive to part of the parameter space of the cNMSSM whereas, trivially, more
luminosity is required in case of heavier squarks and gluinos. In any
case we believe that the cuts proposed here are the most sensitive ones
to the parameter space of the cNMSSM. In addition we have discussed in
how far a refined analysis employing both the signal rate and the
maximum of $M_\mathrm{eff}$ allows to distinguish the cNMSSM
from the MSSM in the stau coannihilation region.

In the near future the LHC is on track to accumulate an integrated luminosity of 1~fb$^{-1}$ 
at 7~TeV c.m. energy at the end of 2011.
 We have estimated the number of signal events for the point P520,
if we lower the cNMSSM cuts correspondingly: for two jets we require
$p_\mathrm{T} > 50$ and 20~GeV, respectively, and
$E_\mathrm{T}^\mathrm{miss} > 200$~GeV. We obtain about 5 signal events
passing these cuts. 

For our analysis of signals of the cNMSSM at 14~TeV c.m. energy we have
left aside the presence of two soft $\tau$ leptons per event which
represent, in principle, a spectacular signature for this class of
models. In some regions of the parameter space of the cNMSSM -- for very
small values of $\lambda$ and/or a small $\widetilde{\tau}_1$-$\chi_1^0$
mass difference -- the life time of $\widetilde{\tau}_1$ can be very
small leading to displaced vertices of the decay $\widetilde{\tau}_1 \to
\tau + \chi_1^0$ into these soft $\tau$-leptons. Using dedicated
track-based algorithms, the search for these soft $\tau$-leptons
originating from displaced vertices is perhaps not completely hopeless.

In the framework of the general NMSSM, a singlino-like LSP can be
accompagnied by a NLSP different from $\widetilde{\tau}_1$. The
signatures of these scenarios would be very different from the ones
discussed here (depending on the nature of the NLSP), and should also be
investigated in the future.

\section*{Acknowledgments}

We would like to thank T. Plehn and P. Wienemann for numerous
discussions, K. Mawatari for help in interfacing PYTHIA with TAUOLA for
supersymmetric models, and Sebastien Binet (LAL) for his invaluable help
in automizing the production of the Alpgen samples. U.E. wishes to thank
the Institut f\"ur Theoretische Physik in Heidelberg, where this work
was started, for hospitality.


\end{document}